\documentclass[%
 aip,
 amsmath,amssymb,
 reprint,%
]{revtex4-1}

\usepackage[utf8]{inputenc}
\usepackage{graphicx}
\usepackage{dcolumn}
\usepackage{bm}
\usepackage [autostyle, english = american]{csquotes}
\MakeOuterQuote{"}
\usepackage[utf8]{inputenc}
\usepackage[T1]{fontenc}
\usepackage{mathptmx}
\usepackage{etoolbox}
\usepackage{accents}
\usepackage{colortbl}
\usepackage{enumitem}
\usepackage{comment}
\usepackage{array,multirow}
\usepackage{float}

\newcommand{\etal}{\textit{et al.}}

\newcommand{\SI}{{Supporting Information }}

\newcommand{\sitab}[1]{{Tab.~#1 in the Supporting Information}}
\newcommand{\simulttab}[2]{{TABLES #1 and #2 in the Supporting Information}}
\newcommand{\sifig}[1]{{Fig.~#1 in the Supporting Information}}
\newcommand{\kcalmol}{\mbox{kcal$\cdot$mol$^{-1}$}}

\makeatletter
\def\@email#1#2{%
 \endgroup
 \patchcmd{\titleblock@produce}
  {\frontmatter@RRAPformat}
  {\frontmatter@RRAPformat{\produce@RRAP{*#1\href{mailto:#2}{#2}}}\frontmatter@RRAPformat}
  {}{}
}%
\makeatother
\begin{document}

\preprint{AIP/123-QED}

\title[Modified Opposite-Spin-Scaled Double-Hybrid Functionals]{Modified Opposite-Spin-Scaled Double-Hybrid Functionals}

\author{Golokesh Santra*}
    \altaffiliation{these authors contributed equally}
    \email{santra@kofo.mpg.de}    
    \affiliation{Max-Planck-Institut für Kohlenforschung, Kaiser-Wilhelm-Platz 1, 45470 Mülheim an der Ruhr, Germany}
\author{Markus Bursch*}
    \altaffiliation{these authors contributed equally}
    \email{bursch@thch.uni-bonn.de}
    \affiliation{FACCTs GmbH, 50677, Köln, Germany}
        \affiliation{Mulliken Center for Theoretical Chemistry, Universität Bonn, Beringstr. 4, 53115 Bonn, Germany}
\author{Lukas Wittmann*}
    \altaffiliation{these authors contributed equally}
    \affiliation{Mulliken Center for Theoretical Chemistry, Universität Bonn, Beringstr. 4, 53115 Bonn, Germany}
    \email{wittmann@thch.uni-bonn.de}   
\date{\today}

\begin{abstract}
We investigate the potential performance improvements of double-hybrid density functionals by replacing the standard opposite-spin-scaled MP2 (SOS-MP2) with the modified opposite-spin-scaled MP2 (MOS-MP2) in the nonlocal correlation component.
Using the large and diverse GMTKN55 dataset, we find that MOS-double hybrids provide significantly better accuracy compared to SOS-MP2-based double hybrids when empirical dispersion correction is not employed.
The non-covalent interaction subsets account for the majority of this improvement.
Adding the DFT-D4 dispersion correction to MOS-type double hybrids does not provide any superior performance over conventional dispersion-corrected SOS-MP2-based double hybrids.
Nevertheless, for nine tested transition metal sets, dispersion-corrected spin-component-scaled (SCS) double hybrids are still significantly better than any MOS-double hybrid functional.
 
\end{abstract}

\maketitle

\section{\label{sec:intro}Introduction}

Kohn–Sham density functional theory (DFT) is one of the main cornerstones of modern computational chemistry, valued for its ability to achieve a good balance between accuracy and computational efficiency, especially with increasing system size.\cite{Teale2022,bestpractice}
Perdew's \textit{Jacob's Ladder}\cite{jacobsladder}  categorizes density functionals based on the type of density or orbital information they utilize within the approximation.
Each ascending rung on the ladder is believed to offer greater accuracy than the one before it, though this improvement typically comes with increased computational cost.
The highest, fifth rung, is represented by the so-called double-hybrid (DH) functionals that include both, an admixture of Hartree--Fock exchange (HFx) and a wavefunction theory-based correlation contribution into the density functional formulation.
The latter is usually computed by second-order perturbation theory (PT2), with second-order M{\o}ller-Plesset theory (MP2) being the most prominent method of choice.\cite{b2plyp,Goerigk2014,Mehta2018,Martin2019,dh_xu,dh23}

Despite the popularity of MP2 as a component of DHs, it suffers from systematic problems, such as its divergent behavior for small orbital energy gaps and the systematic underestimation of long-range dispersion interactions.

The first problem can be addressed by replacing simple MP2 with the direct random phase approximation\cite{dRPA} (dRPA) correlation or by regularizing the MP2 energy expression, e.g., using the $\kappa$-regularization scheme proposed by Shee \etal\cite{Shee2021_reg_mp2}
In a recent study, we showed that dispersion-corrected dRPA-based double hybrids are only as good as MP2-based double hybrids for main group chemistry problems, while dRPA-based double hybrids perform significantly better for metal-organic barrier heights.\cite{santra-dRPA}
The second possibility has also been investigated for the double-hybrid functionals.\cite{santra-kdh,r2SCAN-D4-dh}
Martin and coworkers have shown that for typical double hybrids, which usually employ a large percentage ($\sim$70\% or higher) of HF-exchange, using $\kappa$-MP2 correlation instead of canonical MP2 has no extra benefit.\cite{santra-kdh}
The regularized DHs are better than their standard counterparts, only if a smaller percentage of HF-exchange ($\sim$50\%) is used without the spin-component scaling of the MP2 energy.
In a more recent study, \citeauthor{r2SCAN-D4-dh} have found that $\kappa$Pr\textsuperscript{2}SCAN50, employing only 50\% HFx, performs marginally better overall than the regular Pr\textsuperscript{2}SCAN50, while improving performance for small orbital-energy gap systems and thus adding a layer of robustness for such systems.\cite{r2SCAN-D4-dh}

The second problem is often addressed by adding a dispersion correction to the total electronic energy.
This is a common issue with many density functional approximations (DFAs), as they often fail to accurately describe long-range correlation effects, leading to a systematic underestimation of London dispersion interactions.
As double hybrids usually contain a sealed-down portion of MP2 correlation, they also suffer from a similar issue; therefore, additional London dispersion corrections are required.
Some popular dispersion correction schemes include Grimme's D3\cite{dftd3, dftd3bj} and D4\cite{dftd4,d4coeff,d4periodic,dftd4-ac} corrections, different variants of Vydrov and van Voorhis's VV10 model,\cite{vv10,nl_a, nl_b} the exchange-hole dipole moment (XDM) model of Becke and Johnson,\cite{xdm1,xdm2,xdm3,xdm4} or the Tkatchenko--Scheffler (TS) method.\cite{ts-vdw-1,ts-vdw-2}
Specifically, the efficient DFT-D approach has proven reliable in countless quantum chemical applications and workflows.\cite{accounts,grimme2021,Nicolas2022,Grimme2024}

Even though both spin-component-scaled (SCS) MP2 and spin-opposite-scaled (SOS) MP2 provide a reasonable result within short and medium ranges, they fail to correctly describe the long-range correlation.
To address this issue, Lochan, Jung, and Head-Gordon proposed a distance-dependent modification, where they split the electron interaction operator ($1/r_{12}$) of SOS-MP2 into a short and a long-range part.\cite{MOS-MP2}
In the short-range regime, the MP2 energy is scaled by the usual factor SOS-MP2 of 1.3, but the long-range scaling factor is adjusted to 2.0 -- which recovers the asymptotic MP2 interaction energy of distant fragments.
This modified opposite-spin MP2 (MOS-MP2) has been shown to provide a significant improvement over SOS-MP2 for a variety of chemical problems involving both short-range and long-range interactions.\cite{MOS-MP2}
In this work of \citeauthor{MOS-MP2}, it was already mentioned, that an extension to the double-hybrid scheme could benefit the insufficient description of long-range dispersion in double-hybrid density functional theory.

The main objective of the present study is to demonstrate that an extension of the double-hybrid scheme to MOS-MP2 is possible and delivers good performance to possibly promote further developments in this field.
Moreover, we shall investigate whether the modified opposite-spin-scaled double hybrids outperform their corresponding SOS-MP2-based counterparts when empirical dispersion correction is added.

\section{\label{sec:theory}Theory}

\subsection{MOS-MP2}

It is well known that SOS-MP2 systematically underestimates long-range correlation interactions.
This can be accounted for by making the scaling factor distance-dependent, resulting in a range-separation analogous to range-separation in Hartree--Fock exchange, but for the MP2 correlation part.\cite{MOS-MP2}
The exchange operator in the MP2 integrals is replaced by the MOS operator
\begin{equation}\label{eq:mos}
    \widehat{g}_\omega(\textbf{r})=\dfrac{1}{\textbf{r}}+c_\mathrm{MOS}\dfrac{\mathrm{erf}(\omega \textbf{r})}{\textbf{r}},
\end{equation}
where $\omega$ determines the strength of attenuation.
It leads to the modified integral
\begin{equation}
    \tilde{I}_{ia}^{jb} =  \int d\textbf{r} \int d\textbf{r}\;' \phi_i(\textbf{r}) \phi_a(\textbf{r}) \widehat{g}_\omega(\textbf{r}-\textbf{r}\;') \phi_j(\textbf{r}\;') \phi_b(\textbf{r}\;');
\end{equation}
which provides the following energy expression for MOS-MP2 
\begin{equation}\label{eq:mos-mp2}
    E_\mathrm{MP2}^\mathrm{MOS} = - \sum_{ia}^{\alpha} \sum_{jb}^{\beta} \dfrac{\tilde{I}_{ia}^{jb}(\omega) \tilde{I}_{ia}^{jb}(\omega)}{[\epsilon_a + \epsilon_b - \epsilon_i - \epsilon_j]},
\end{equation}
where $i$, $j$ are the occupied and $a$, $b$ are the virtual orbitals with corresponding eigenvalues, $\epsilon$.
The MOS operator, $\widehat{g}_\omega(\textbf{r})$, (Eq.~\ref{eq:mos}) depends on two parameters, $\omega$ and $c_\mathrm{MOS}$. 
$c_\mathrm{MOS}$ can be fixed at \(\sqrt{2}-1\) by using the condition of
\begin{equation}
    \lim_{\mathbf{r}\to\infty}E_\mathrm{MP2}^\mathrm{MOS}=2E_\mathrm{MP2}^\mathrm{OS}.
\end{equation}
Hence, our MOS-MP2 scheme depends only on a single parameter, $\omega$.
For main group chemistry problems, the \citeauthor{MOS-MP2} recommends $\omega=0.6$.
Similar to SOS-MP2,\cite{sosmp2} for MOS-MP2, the scaling factor for the same-spin MP2 correlation energy components is 0, but the parameter for the opposite-spin MP2 correlation ranges from 1.3 to 2.0. 

\subsection{Double Hybrids: Hartree--Fock and MP2 Admixture}

In 2006, \citeauthor{grimme2006} proposed the so-called \textit{double hybrid} functionals by combining a fraction of exact exchange and nonlocal GLPT2\cite{GLPT2} (second-order Görling–Levy perturbation theory) correlation with the semilocal DFT exchange and correlation components.\cite{grimme2006}
These functionals have the following expression for the exchange-correlation energy
\begin{equation}\label{eq:dh}
    E_{XC}^\mathrm{DH} = a_{X}E_{X}^\mathrm{HF} + (1-a_{X})E_{X}^\mathrm{DFA}+ (1-a_{C})E_{C}^\mathrm{DFA} + a_{C}E_{C}^\mathrm{MP2},
\end{equation}
where $E_{X}^\mathrm{DFA}$ and $E_{C}^\mathrm{DFA}$ represent the semilocal exchange and correlation energy components;
$E_{X}^\mathrm{HF}$ and $E_{C}^\mathrm{MP2}$ are the HF-exchange and GLPT2 correlation energies -- {$a_{X}$} and {$a_{C}$} are the respective parameters.
Previously, the term \textit{gDH} was used for these types of functionals.\cite{santra-wDH}
Later, Martin and coworkers showed that using separate parameters for the same and opposite-spin MP2 correlation (i.e., $a_{OS}$ and $a_{SS}$) improved the accuracy of DHs for main-group thermochemistry and harmonic frequencies.\cite{Kozuch2010,kozuch2011,kozuch2013,manoj-hfreq,santra2019,dh-f12-freq}

Another family of double hybrids, which are often referred to as xDHs, uses full semilocal correlation instead for the generation of KS reference orbitals.\cite{Manoj_martin_xDH, santra-wDH, santra-xyg9}
It was argued that such orbitals are more appropriate as a basis for GLPT2 than the damped-correlation orbitals in the gDHs.\cite{xu-xyg09, xu-xyg12}
However, this argument has been refuted on empirical grounds by \citeauthor{grimme-gmtkn30} and by \citeauthor{Manoj_martin_xDH}.\cite{Manoj_martin_xDH, grimme-gmtkn30} 
The XYG-family of double hybrids from \citeauthor{xu-xyg09}, the xDSD(xDOD) and the XYG8 functionals by \citeauthor{santra-xyg9} belong to this category.\cite{xu-xyg09, xu11, xu-xyg12,xu13,xu21,santra-xyg9,Manoj_martin_xDH, santra-wDH, santra-xDH-correction}

The exchange-correlation energy for a modified opposite-spin scaled double hybrid (MOS-DH) functional is expressed as  
\begin{equation}\label{eq:mos-dh}
    \begin{split}
    E_{XC}^\mathrm{MOS-DH} = a_{X}E_{X}^\mathrm{HF} + (1-a_{X})E_{X}^\mathrm{DFA} \\
    + a_{C,\mathrm{DFA}}E_{C}^\mathrm{DFA} + a_{OS}E_{C}^\mathrm{MOS-MP2},
    \end{split}
\end{equation}
where $E_{X}^\mathrm{DFA}$, $E_{X}^\mathrm{HF}$, and $E_{C}^\mathrm{DFA}$ represent the same energy component as in Eq.~\eqref{eq:dh}.
$a_{X}$ and $a_{C,\mathrm{DFA}}$ are the parameters for the HF-exchange and the semilocal-correlation energy components. 
$E_{C}^\mathrm{MOS-MP2}$ is the MOS-MP2 correlation energy component, and $a_{OS}$ is the corresponding parameter.
We refer to these new functionals as MOS$_n$-XC in the remaining text, where XC is a combination of DFA exchange and correlation and $n$ is the percentage of HF-exchange (i.e., $n=100a_X$). In passing, we must note that $a_{C,\mathrm{DFA}}$ in Eq.~\eqref{eq:mos-dh} and ($1-a_C$) in Eq.~\eqref{eq:dh} are the same parameters.

On the other hand, for the modified opposite-spin scaled version of the XYG8-family functional, the exchange-correlation energy expression is
\begin{equation} \label{eq:mos-xyg}
    \begin{aligned}
    E_{XC}^\mathrm{MOS-XYG@B\textit{m}LYP} = & a_{X}E_{X}^\mathrm{HF}+a_{X,\mathrm{DFA}}E_{X}^\mathrm{B88}+a_{X,\mathrm{S}}E_{X}^\mathrm{Slater} \\
    &+ a_{C,LDA}E_{C}^\mathrm{VWN5} + a_{C,\mathrm{DFA}}E_{C}^\mathrm{LYP} \\
    &+ a_{OS}E_{C}^\mathrm{MOS-MP2},
    \end{aligned}
\end{equation}
where $E_{X}^\mathrm{HF}$, $E_{C}^\mathrm{MOS-MP2}$, $a_{X}$, and $a_{OS}$ represent the same as in Eq.~(\ref{eq:mos-dh});
$E_{X}^\mathrm{B88}$ and $E_{X}^\mathrm{Slater}$ are the exchange energy components from the Becke88 generalized gradient approximation (GGA) and Slater-type local density approximation (LDA);
$a_{X,\mathrm{DFA}}$ and $a_{X,\mathrm{S}}$ are corresponding parameters, respectively.
The semilocal GGA and LDA correlation energies are represented by $E_{C}^\mathrm{LYP}$ and $E_{C}^\mathrm{VWN5}$, where $a_{C,\mathrm{DFA}}$ and $a_{C,\mathrm{LDA}}$ are the respective coefficients for those energy components.
The reference orbitals used in MOS-XYG are B\textit{m}LYP [i.e., \textit{m}\% HFx + ($100-m$)\% DFTx + 19\% VWN5 + 81\% LYP].
We employed the same GGA-type and LDA-type exchange and correlation during the orbital generation as well as the final energy evaluation step.

\subsection{\label{sec:disp}DFT-D4 Dispersion Correction}

The default atomic-charge dependent D4 dispersion correction including Axilrod-Teller-Muto\cite{ATM1,ATM2} (ATM) type three-body contributions was applied according to Eq.~(\ref{eq:d4}) with atomic indices $A$, $B$, and $C$, their distance $R_{AB}$, the $n$\textsuperscript{th} dispersion coefficient $C^{AB}_{(n)}$, and the angle-dependent term $\theta_{ABC}$
\begin{subequations}
    \begin{align}
        E^{D4}_{disp} = &-\frac{1}{2}\sum_{AB}\sum_{n=6,8} s_{n}\frac{C^{AB}_{(n)}}{{R^{(n)}_{AB}}} f^{(n)}_{damp}(R_{AB}) \\
        &-\frac{1}{6}\sum_{ABC} s_{9}\frac{C^{ABC}_{(9)}}{{R^{(9)}_{ABC}}} f^{(9)}_{damp}(R_{ABC},\theta_{ABC})
    \end{align}
    \label{eq:d4}
\end{subequations}
where $f^{(n)}_{BJ}(R_{AB})$ corresponds to the default Becke--Johnson~(BJ) damping function\cite{dftd3bj} according to Eq.~\eqref{eq:bj}:
\begin{equation}
 f^{(n)}_{BJ}(R_{AB})=\frac{R_{AB}^{(n)}}{R_{AB}^{(n)}+(a_{1}R_{0}^{AB}+a_{2})^{(n)}}\,.
\label{eq:bj}
\end{equation}
The usually fitted parameters for a non-DH functional are $s_8$, $a_1$, and $a_2$. 
For a DH, $s_6$ must also be adjusted due to the presence of the MP2 correlation term. 
As in Refs.~\citenum{santra-wDH, santra-xDH-correction, r2SCAN-D4-dh}, the parameter controlling the three-body contribution was set to $s_9=1$ for all MOS-MP2-based double hybrids.

\section{\label{sec:methods}Computational Details}

Unless otherwise specified, all calculations were performed using the Q-Chem 5.4 quantum chemistry program package.\cite{qchem5}
The Weigend-Ahlrichs quadruple-$\zeta$ basis set def2-QZVPP\cite{def2basen} was used for all calculations.
For seven GMTKN55 subsets (WATER27,\cite{water} RG18,\cite{gmtkn} IL16,\cite{il16} G21EA,\cite{g21} AHB21,\cite{il16} BH76,\cite{bh76_1,bh76_2,bh76_3} and BH76RC\cite{bh76_3}) the diffuse-function-augmented def2-QZVPPD\cite{def2diffuse} was employed instead.
The matching effective core potentials (ECPs)\cite{def2ecp, def2ecp_2} for heavy elements with Z > 36 were generally employed.
For the MP2 part, RI approximation was applied to accelerate the calculations in conjunction with the def2-QZVPPD-RI\cite{def2-RI-Hatting, def2-RI-Rappoport} auxiliary basis set.
The SG-3\cite{SG-3grid} integration grid was employed, except for the SCAN (strongly constrained and appropriately normed) variants,\cite{scan2015} where an unpruned (150, 590) grid was used for its severe integration grid sensitivity.\cite{grid_dep_SCAN}
Similar to Ref.~\citenum{santra-xyg9}, for the SCF energy components of MOS-XYG double hybrids, the MRCC2020 package was used.\cite{kallay2020mrcc}

DFT-D4 dispersion corrections were calculated with the \texttt{dftd4} 3.4.0 standalone program.\cite{dftd4,d4coeff,d4periodic,dftd4-ac}

Reference geometries for the GMTKN55 benchmark sets were taken from Ref.~\citenum{gmtkn}.
Additionally, the transition-metal chemistry sets CUAGAU-2,\cite{CUAGAU2} LTMBH,\cite{LTMBH} MOBH35,\cite{MOBH35,MOBH35err,Dohm2020} MOR41,\cite{mor41} ROST61,\cite{rost61} TMBH,\cite{TMBHa,TMBHb,TMBHc,TMBHd} TMCONF16,\cite{TMCONF16} TMIP,\cite{TMIP} and WCCR10\cite{WCCR10,WCCR10err} are evaluated -- this compilation of benchmark sets will be referred to as \textit{TM} throughout this document. 

\section{\label{sec:param}Parametrization Strategy}

The modified opposite-spin scaled double hybrid functionals have been parameterized using the GMTKN55 benchmark suite.\cite{gmtkn}
This dataset consists of 55 types of chemical problems, which can be further divided into five subsets: basic thermochemistry of small molecules, barrier heights, large molecule reactions, intermolecular interactions, and conformer energies.

Originally proposed by \citeauthor{gmtkn}, the WTMAD-2\cite{gmtkn} (weighted total mean absolute deviation) has been used as the primary metric for the performance evaluation and parameter optimization of the MOS-DHs.
From a statistical viewpoint, MAD (mean absolute deviation) is a more robust metric than RMSD (root-mean-square deviation), as the former is more resilient to a small number of large outliers than the latter.\cite{huber2004robust}
For a normal distribution without a systematic error, RMSD $\approx \frac{5}{4} \mathrm{MAD}$.\cite{geary1935ratio}
See Appendix B in the \SI for the definition of the used statistical measures.  

The constructed MOS-DH functionals have four empirical parameters: 
\begin{enumerate}[label=(\alph*)]
    \item fraction of exact exchange ($a_X$)
    \item fraction of the semilocal DFT correlation ($a_{C,\mathrm{DFA}}$)
    \item coefficient for the MOS-MP2 correlation ($a_{OS}$)
    \item parameter $\omega$ controls the level of attenuation of the MOS operator
\end{enumerate}

The MOS-XYG double hybrid scheme contains three additional parameters: 
\begin{enumerate} [label=(\alph*),resume]
    \item fraction of B88 exchange ($a_{X,\mathrm{DFA}}$)
    \item fraction of Slater exchange ($a_{X,\mathrm{S}}$)
    \item fraction of local correlation ($a_{C,\mathrm{LDA}}$)
\end{enumerate}

Powell’s BOBYQA (bound optimization by quadratic approximation) derivative-free constrained optimizer was used for the optimization.\cite{powell2009bobyqa}
For a given set of \{$a_X$, $a_{C,\mathrm{DFA}}$, $\omega$\}, it is possible to obtain the optimal value of {$a_{OS}$} without any further electronic structure calculations simply by extracting individual energy components from the calculations, evaluating total energies and hence WTMAD-2 for a given {$a_{OS}$}, and minimizing WTMAD-2 with respect to {$a_{OS}$} using BOBYQA.
It can be considered as the \textit{microiteration} loop. In comparison, the outer \textit{macroiteration} loop consists of varying \{{$a_X$}, {$a_{C,\mathrm{DFA}}$}, $\omega$\} and reevaluating the full GMTKN55 using the updated set of parameters. For the revised DSD and DOD functionals, we found that {$a_{C,\mathrm{DFA}}$} could be safely included in the microiteration, but {$a_X$} could not due to its strong coupling with the MP2 scaling factors.\cite{santra2019}
Hence, we have adopted the practice of microiterating \{{$a_{C,\mathrm{DFA}}$}, {$a_{OS}$}\} at every macroiteration using BOBYQA.
We must note that, with full microiteration cycles, additional macroiterations beyond the first typically do not have significantly improved performance unless the starting guess is especially poor. Hence, the output of the first cycle is reported.
The optimum value of $\omega$\ for each {$a_X$} was determined manually by interpolation.

For each exchange-correlation combination, the above-mentioned process is repeated with multiple {$a_X$} values to find the best MOS-DH, which offers the lowest WTMAD-2.

While using D4 dispersion correction with MOS-DHs, four extra parameters needed to be included in the microiteration loop: s$_6$, s$_8$, s$_9$, $a_1$, and $a_2$. Hence, for each \{{$a_X$}, $\omega$\} set, we optimized \{{$a_{C,\mathrm{DFA}}$}, {$a_{OS}$}, s$_6$, s$_8$, s$_9$, $a_1$, $a_2$\}. Like in revDSD-PBEP86-D4 and xDSD$_{75}$-PBEP86-D4, the $s_9=1$ constraint was used across the board while optimizing the other microiteration parameters of MOS-DH-D4 functionals. 

For optimizing the MOS-XYG parameters, the same protocol as in Ref.~\citenum{santra-xyg9} was used.
Except for $\omega$, all the parameters were included in the microiteration loop.
For the B\textit{m}LYP reference orbitals, the value of $m$ has been optimized manually by interpolation.

\section{\label{sec:discussion}Results and Discussion}

\subsection{Main Group Chemistry Problems (GMTKN55)}

\begin{figure}[!htbp]
    \centering
    \includegraphics[width=8.8cm]{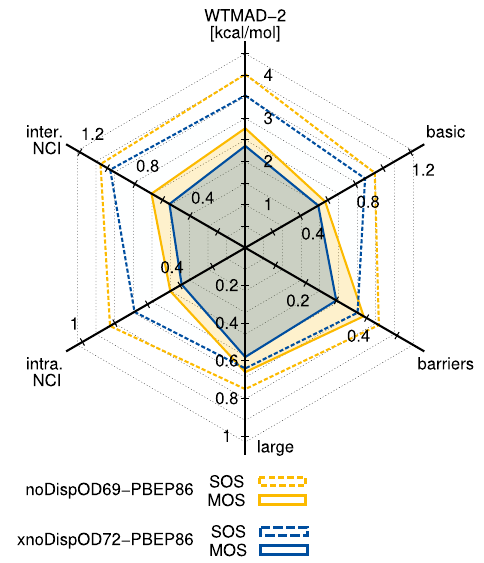}
    \caption{Effect of using MOS-MP2 in a double hybrid functional on total WTMAD-2\textsubscript{GMTKN55} and $\Delta$WTMAD-2 contributions from five major subsets of GMTKN55. The numbers are given in \kcalmol.}
    \label{fig:mos-effect}
\end{figure}

To evaluate the impact of using MOS-MP2 as the nonlocal correlation part of double hybrids, the SOS-MP2 components of noDispOD$_{69}$-PBEP86 and xnoDispOD$_{72}$-PBEP86 are replaced with MOS-MP2, and the corresponding parameters are reoptimized, keeping the other parameters unchanged. 
As a result, the WTMAD-2\textsubscript{GMTKN55} values improved by 1.25 and 1.17 \kcalmol\ for their respective MOS-MP2-based counterparts.
For both cases, the lion’s share of this improvement comes from basic thermochemistry and the two non-covalent interaction subsets (Fig.~\ref{fig:mos-effect}).
Expecially the W4-11,\cite{w4-11} PCONF21,\cite{pconf_1,pconf_2} and S66\cite{s66_orig} subsets of the GMTKN55 improved significantly. 
Because MOS-MP2 uses a distance-dependent scaling factor (ranging from 1.3 to 2.0), the optimized $a_{OS}$ for the MOS-MP2-based double hybrid is smaller than its SOS-MP2-based counterpart noDispOD$_{69}$-PBEP86.
The ratio of $a_{OS}$ between SOS-MP2 and MOS-MP2-based double hybrids is 1.4, which increases to 1.7 with the increase of $a_X$ from 0.69 to 0.82 (\sitab{S2}).

Using MOS-MP2 instead of SOS-MP2 greatly enhances the accuracy of dispersion-uncorrected double hybrids, prompting us to optimize all parameters of MOS-DH functionals further. The final parameters and WTMAD-2\textsubscript{GMTKN55} values for the GMTKN55 for various MOS-DHs and their corresponding SCS-MP2-based dispersion-free counterparts are listed in Tab.~\ref{tab:wtamd2_and_params}.

\begin{table*}[!htbp]
\centering
\caption{Final parameters and total WTMAD-2\textsubscript{GMTKN55} of the MOS-DHs and their respective spin-component-scaled, dispersion uncorrected DHs on the GMTKN55 in \kcalmol.}
\label{tab:wtamd2_and_params}
\begin{ruledtabular}
\renewcommand{\arraystretch}{1.3}
\begin{tabular}{lccccccccc}
\textbf{Functional} & \textbf{WTMAD-2}	& \textbf{$\omega$} & \textbf{{$a_{X}$}} & \textbf{$a_{X,\mathrm{DFA}}$} & \textbf{$a_{X,S}$} & \textbf{{$a_{C,\mathrm{DFA}}$}}	& \textbf{$a_{C,LDA}$}	& \textbf{{$a_{OS}$}} & \textbf{$a_{SS}$} \\
\hline
MOS$_{76}$-PBEP86    & 2.48 & 0.5000 & 0.7600 & 0.2400    & --- & 0.4371 & --- & 0.5602 & [0] \\
xMOS$_{78}$-PBEP86  & 2.27  & 0.6500 & 0.7800 & 0.2200    & --- & 0.4056 & --- & 0.5373 & [0] \\
MOS-XYG@B$_{50}$LYP & 2.05  & 0.9000 & 0.8871 & --0.1087 & 0.2165 & [0] & 0.3982 & 0.4406 & [0] \\
\hline
noDispSD$_{82}$-PBEP86    & 2.89 & --- & 0.8200 & 0.1800    & --- & 0.3073 & --- & 0.7426	& 0.3782 \\
xnoDispSD$_{82}$-PBEP86  & 2.51  & --- & 0.8200 & 0.1800    & --- & 0.2797 & --- & 0.7678	& 0.3521 \\
XYG8[$f_2$]@B$_{25}$LYP\cite{santra-xyg9} & 1.92  & --- & 0.9144 & --0.1537  & 0.2132 & [0] & 0.4928 & 0.4540 & 0.2780 \\
\end{tabular}
\end{ruledtabular}
\end{table*}

\begin{figure*}[!htbp]
    \centering
    \includegraphics[width=\textwidth]{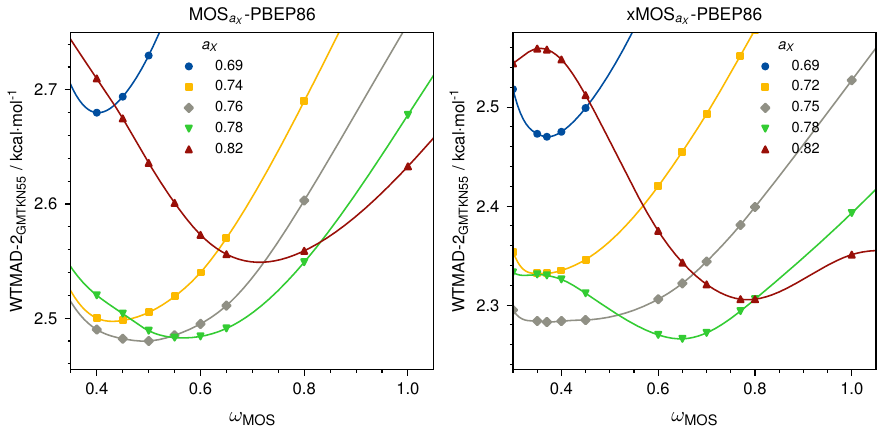}
    \caption{Dependence of WTMAD-2\textsubscript{GMTKN55}\textsubscript{GMTKN55} (\kcalmol) on the MOS-MP2 range-separation parameter $\omega$ for MOS$_{a_X}$-PBEP86 and xMOS$_{a_X}$-PBEP86.}
    \label{fig:mos-xmos-fig}
\end{figure*}

While using the PBE-P86 exchange-correlation combination, varying the percentage of HF-exchange and the MOS-MP2 rang-separation parameter ($\omega$) simultaneously, we obtained the lowest WTMAD-2\textsubscript{GMTKN55} of 2.48~\kcalmol\ with $a_X=0.76$ and $\omega=0.50$ (Fig.~\ref{fig:mos-xmos-fig} and Tab.~\ref{tab:wtamd2_and_params}).
Among the five major subcategories of the GMTKN55, barrier heights and non-covalent interactions benefit the most upon using MOS-MP2 (see Tab.~\ref{tab:subset_wtamd2}).
The major share of this improvement comes from the BH76,\cite{bh76_1,bh76_2,bh76_3} S66,\cite{s66_orig} RSE43,\cite{rse43} and PCONF21\cite{pconf_1,pconf_2} subsets.
On the other hand, the performance on, e.g., the bond-separation reactions of saturated hydrocarbons subset BSR36 is worsened.\cite{bsr36_1,bsr36_2}
When compared to the revDOD-PBEP86-D4, the absence of empirical dispersion correction significantly deteriorates the accuracy of MOS$_{76}$-PBEP86 for the pericyclic reaction barrier heights,\cite{bh76_3,bhperi_1,bhperi_2,bhperi_3} BH76,\cite{bh76_1,bh76_2,bh76_3} RSE43,\cite{rse43} and RG18\cite{gmtkn} subsets, but outperforms revDOD-PBEP86-D4 for the tautomer relative energies.

\begin{figure*}[!htbp]
    \centering
    \includegraphics[width=\textwidth]{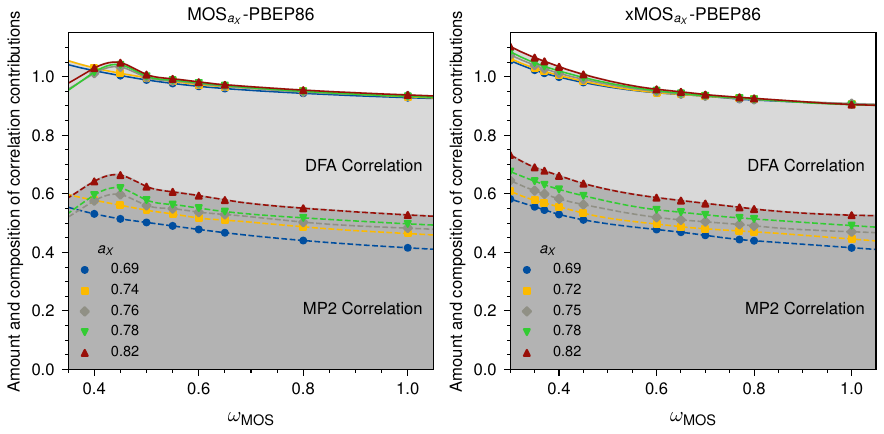}
    \caption{Dependence of the optimised amounts of MOS-MP2 and DFA correlation on the MOS-MP2 range-separation parameter $\omega$ for MOS$_{a_X}$-PBEP86 and xMOS$_{a_X}$-PBEP86.}
    \label{fig:dfa and mp2 corr}
\end{figure*}

For a specific MOS-MP2 range-separation parameter $\omega$, an increasing percentage of HF-exchange yields a larger optimized value of the MOS-MP2 correlation parameter {$a_{OS}$}, while the fraction of the semilocal DFT correlation decreases (Fig.~\ref{fig:dfa and mp2 corr}).
On the other hand, for a fixed value of {$a_X$}, increasing $\omega$ yields a lower {$a_{OS}$}.
For a very small $\omega$ (i.e., $\omega<0.5$) {$a_{C,\mathrm{DFA}}$} decreases (\sitab{S9}).
The trend of the WTMAD-2\textsubscript{GMTKN55} against the optimized $a_{OS}$ is shown in \sifig{S2}.  

For the revDSD- and revDOD-family double hybrids, we achieved the lowest WTMAD-2\textsubscript{GMTKN55} using only the PBE-P86 exchange-correlation combination, as noted by \citeauthor{santra2019}.
Until now, we have only used the PBE exchange and P86 correlation.
Similarly, however, for the MOS-DHs, using the SCAN-SCAN and PBE-PBE exchange-correlation combinations resulted in worse performance compared to their PBE-P86-based counterparts.
Among the PBE-based functionals, the lowest WTMAD-2\textsubscript{GMTKN55} was obtained with $a_X=0.78$ and $\omega=0.50$.
On the other hand, for the MOS$_n$-SCAN type functionals, the optimum values of the parameters are $0.74$ and $0.90$, respectively (Tab.~S2 and Fig.~S1 in the \SI).
For each \{{$a_X$}, $\omega$\} combination, the WTMAD-2\textsubscript{GMTKN55} error statistics for the SCAN- and PBE-based MOS-DHs are listed in \simulttab{S6}{S7}, respectively. 
        
For the xDH variant, we obtained the lowest WTMAD-2\textsubscript{GMTKN55} of $2.27$ \kcalmol\ by employing $a_X=0.78$ and $\omega=0.65$ (Fig.~\ref{fig:mos-xmos-fig}).
The improvement in WTMAD-2\textsubscript{GMTKN55} for the MOS-type double hybrid is $0.21$, whereas, for the noDispSD functionals, that improvement is $0.37$ \kcalmol\ (Tab.~\ref{tab:wtamd2_and_params}).
The majority of that improvement for MOS-DHs originates from the large-species reaction energies.
Further inspection reveals that for the RSE43 set, the performance of xMOS$_{78}$-PBEP86 is noticeably better than MOS$_{76}$-PBEP86.
For a fixed value of $\omega$, an increasing amount of HF exchange also leads to a larger optimal value of {$a_{OS}$} and simultaneously smaller {$a_{C,\mathrm{DFA}}$} (Fig.~\ref{fig:dfa and mp2 corr} and \sitab{S10}).
The performance comparison with optimized $a_{OS}$ is given in \sifig{S2}.

\begin{table*}[!htbp]
\centering
\caption{WTMAD-2 and the $\Delta$WTMAD-2 contributions of the five major subsets to total WTMAD-2\textsubscript{GMTKN55} in \kcalmol.}
\begin{ruledtabular}
\renewcommand{\arraystretch}{1.3}
\begin{tabular}{lcccccc}
\textbf{Functional} &\textbf{WTMAD2} & \textbf{basic}	& \textbf{barrier}	& \textbf{large} & \textbf{intra. NCI}  & \textbf{inter. NCI}  \\
\hline
MOS$_{76}$-PBEP86		& 2.48 &	0.52	&	0.37	&	0.64	&	0.41	&	0.55	\\
xMOS$_{78}$-PBEP86		& 2.27	& 0.52	&	0.32	&	0.54	&	0.38	&	0.51	\\
MOS-XYG@B$_{50}$LYP		& 2.05 & 0.52	&	0.29	&	0.43	&	0.37	&	0.44	\\
\hline													
noDispSD$_{82}$-PBEP86		& 2.89 &	0.57	&	0.49	&	0.65	&	0.51	&	0.66	\\
xnoDispSD$_{82}$-PBEP86		& 2.51 &	0.52	&	0.40	&	0.53	&	0.46	&	0.61	\\
XYG8[$f_2$]@B$_{25}$LYP\cite{santra-xyg9}	& 1.92 & 	0.46&	0.23	&	0.36	&	0.43	&	0.44\\
\hline
revDOD-PBEP86-D4\cite{santra-wDH}	& 2.27	& 0.58	& 0.25	& 0.58	& 0.40	& 0.46	\\
xDOD$_{72}$-PBEP86-D4\cite{santra-wDH,santra-xDH-correction}	& 2.20	& 0.57	& 0.23	& 0.51	& 0.41	& 0.47	\\
Pr\textsuperscript{2}SCAN69-D4\cite{r2SCAN-D4-dh} &2.72\textsuperscript{(a)} &0.62 &0.36 &0.58 &0.59 &0.56 \\
\end{tabular}
\begin{footnotesize}

\textsuperscript{(a)}Due to the use of def2-QZVPPD basis set for seven GMTKN55 subsets (WATER27, RG18, IL16, G21EA, AHB21, BH76, and BH76RC) in the present work, the total WTMAD-2\textsubscript{GMTKN55} is 0.09 kcal/mol lower than what is reported in Ref.~\citenum{r2SCAN-D4-dh}
\end{footnotesize}
\end{ruledtabular}
\label{tab:subset_wtamd2}
\end{table*}

Comparing the performance of xMOS$_{78}$-PBEP86 and xnoDispSD$_{82}$-PBEP86, we found that the barrier heights, inter- and intramolecular non-covalent interactions benefit from MOS-MP2 over SCS-MP2.
Although for most of the reaction types of GMTKN55, xMOS$_{78}$-PBEP86 performs better than xnoDispSD$_{82}$-PBEP86, for small molecule atomization energies (W4-11), decomposition energies of a few artificial molecules (MB16-43), and bond-separation reactions of saturated hydrocarbons (BSR36), xnoDispSD$_{82}$-PBEP86 outperforms by a significant margin. 

Interestingly enough, the WTMAD-2\textsubscript{GMTKN55} gap between xMOS$_{78}$-PBEP86 and xDOD$_{72}$-PBEP86-D4 is only $0.07$ \kcalmol.
Except for the basic thermochemistry and intramolecular interactions, xDOD$_{72}$-PBEP86-D4 outperforms xMOS$_{78}$-PBEP86 for the remaining three GMTKN55 subsets (Tab.~\ref{tab:subset_wtamd2} and \sitab{S3}). 

\begin{figure*}[!htbp]
    \centering
    \includegraphics[width=\textwidth]{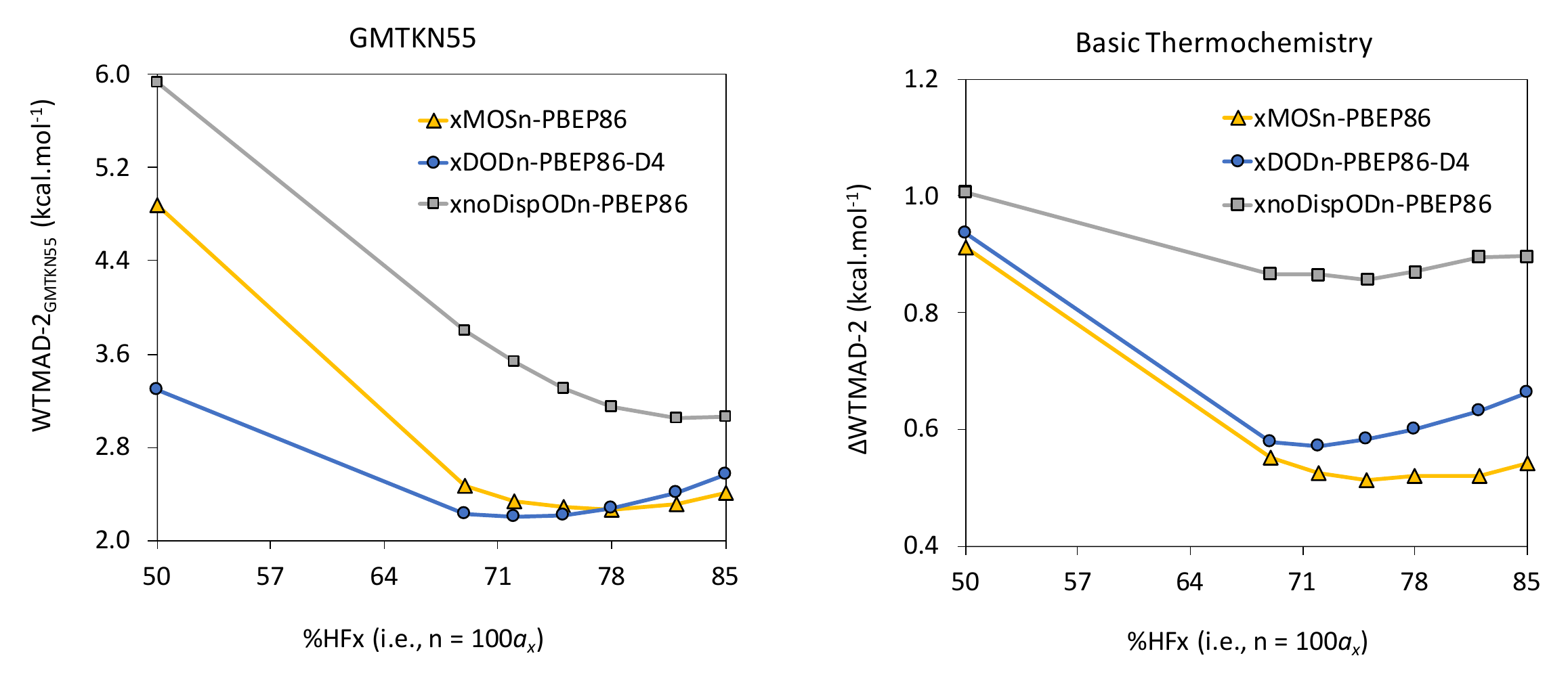}
    \caption{Dependence of total WTMAD-2\textsubscript{GMTKN55} and $\Delta$WTMAD-2 for the basic thermochemistry reactions of GMTKN55 on different percentages of HF-exchange (i.e., $n=100a_X$) in xMOS$_n$-PBEP86, xnoDispSD$_n$-PBEP86, and xDOD$_n$-PBEP86-D4.}
    \label{fig:xmos vs xdod}
\end{figure*}

For a fixed value of $a_X$, the remaining parameters of xMOS$_n$-PBEP86 and xDOD$_n$-PBEP86-D4 are optimized and their performance is evaluated. 
It is clear that the WTMAD-2\textsubscript{GMTKN55} gap decreases gradually from $a_X=0.50$ to $a_X=0.78$.
Beyond 78\% HF-exchange, MOS-DHs outperform the respective SOS-MP2-based dispersion corrected xDHs (Fig.~\ref{fig:xmos vs xdod} and \sitab{S13}).
Only for the basic thermochemistry reactions, xMOS-DHs are better than the respective xDOD-D4 functionals, and performance difference increases with the increasing amount of Hartree--Fock exchange (Fig.~\ref{fig:xmos vs xdod}).
For the remaining four reaction types of the GMTKN55, the xDOD-D4 functionals outperform the xMOS-DHs for a small percentage of HF-exchange.
On the other hand, at larger percentages (e.g., $a_X=0.85$), xMOS-DH offers better performance for barriers compared to the corresponding xDOD-D4.
For inter- and intramolecular non-covalent interactions, both double hybrid schemes offer similar accuracy (\sifig{S3}).
The optimized parameters of xMOS$_n$-PBEP86 and xDOD$_n$-PBEP86-D4 can be found in \simulttab{S11}{S12}.

An increasing amount of Hartree--Fock admixture leads to a decreasing difference in overall main-group performance between xMOS$_n$-PBEP86 and xnoDispOD$_n$-PBEP86.
For basic thermochemistry reactions, the difference between xMOS$_{50}$-PBEP86 and xnoDispOD$_{50}$-PBEP86 is small, but increases at larger percentages of HF-exchange (Fig.~\ref{fig:xmos vs xdod}).
On the other hand, for the remaining four subsets, the $\Delta$WTMAD-2 difference gradually decreases with the increase in $a_X$ (\sifig{S3}).

In a recent study, Martin and co-workers reported that using an elevated percentage of HF-exchange and separate parameters for the local and semilocal exchange and correlation energy components, XYG8[$f_2$]@B$_{25}$LYP offered very good performance (WTMAD-2 of $1.92$ \kcalmol) without employing any empirical dispersion correction.\cite{santra-xyg9}
While other dispersion uncorrected double hybrids listed in Table \ref{tab:wtamd2_and_params} have only 4 empirical parameters, XYG8[$f_2$]@B$_{25}$LYP contains 6.

For the MOS-XYG@B\textit{m}LYP functional, six different types of orbitals were tested, with the percentage of HF-exchange varying from 10\% to 70\%, and for each reference orbital, MOS-MP2 correlation was evaluated with seven different $\omega$ values (\sitab{S7}).
Among the modified SOS-MP2 double hybrids reported in the present study, MOS-XYG@B$_{50}$LYP has the lowest WTMAD-2\textsubscript{GMTKN55} of 2.05 \kcalmol.
Similar to the XYG8 functionals, for MOS-XYG, the optimized value of {$a_{C,\mathrm{DFA}}$} is found to be nearly zero; fixing $a_{C,\mathrm{DFA}} = 0$ during optimization does not influence the overall performance.

Unlike the previous cases, for the XYG family functionals, WTMAD-2\textsubscript{GMTKN55} for the MOS-MP2-based double hybrid is $0.13$~\kcalmol\ higher than its SCS-MP2-based counterpart (Tab.~\ref{tab:wtamd2_and_params}).
Comparing the WTMAD-2\textsubscript{GMTKN55} contributions from Tab.~\ref{tab:subset_wtamd2}, it is evident that the deterioration in the performance of MOS-XYG@B$_{50}$LYP comes from the basic thermochemistry, barrier heights, and large molecule reaction energies.
Upon further inspection, the main cause of the decrease in performance is due to the BH76 and BSR36 subsets.
It is interesting to note that the MOS-MP2 XYG-family double hybrid, MOS-XYG@B$_{50}$LYP, performs marginally better than the dispersion-corrected SOS-MP2-based functional, XYG8[$f_{2os}$]@B$_{30}$LYP.
For the TAUT15, BSR36, and BHPER subsets, the MOS-XYG double hybrid offers better performance than XYG8[$f_{2os}$]@B$_{30}$LYP, but for the BH76, W4-11, BH76RC, and RSE43\cite{rse43} reactions, XYG8[$f_{2os}$]@B$_{30}$LYP is superior.

\subsection{Effect Dispersion Correction}

For this purpose, we use MOS$_n$-PBEP86 and xMOS$_n$-PBEP86 series and combine them with the DFT-D4 dispersion correction.
Contrary to Refs.~\citenum{santra2019,santra-wDH}, the optimized $s_8$ parameters of MOS$_n$-PBEP86 are not always close to zero. For a specific percentage of HF exchange, that parameter only vanishes for small $\omega$ values.
With D4 dispersion correction, the lowest WTMAD-2\textsubscript{GMTKN55}\textsubscript{GMTKN55} was obtained with $a_{X}=0.69$ and $\omega=0.10$.
The WTMAD-2\textsubscript{GMTKN55}s listed in Table S13 indicate that the performance of MOS-PBEP86-D4 is approaching the limit of revDOD-PBEP86-D4.
Conversely, the performance of xMOS$_n$-PBEP86-D4 is somewhat heterogeneous, exhibiting both advantages and disadvantages (\sitab{S14}).
The optimized WTMAD-2\textsubscript{GMTKN55} of xMOS$_{72}$-PBEP86-D4 ($\omega=0.2$) is very close to the xDOD$_{72}$-PBEP86-D4.
Hence, for the GMTKN55 reactions, the D4 dispersion-corrected MOS double hybrids are not better than revDOD-PBEP86-D4 and xDOD$_{72}$-PBEP86-D4.
When dispersion-corrected and uncorrected MOS$_{76}$-PBEP86 ($\omega=0.50$) and xDOD$_{78}$-PBEP86 ($\omega=0.65$) correlation parameters are compared, it is found, that $a_{OS}$ and $a_{C,DFA}$ decrease slightly upon including D4 (\sitab{S15}).  

For the PBEP86, SCAN, and PBE-based MOS-double hybrids, the summation of the empirical {$a_{C,DFA}$} and {$a_{OS}$} are very close to unity.
Setting the constraint of $a_{C,DFA} + a_{OS} = 1.0$ during optimization, parameters and WTMAD-2\textsubscript{GMTKN55} errors as shown in \sitab{S12} are obtained.
We find, that reducing the number of empirically fitted parameters from 4 to 3 has no noticeable influence on the performance of the tested MOS-DHs.
The only exception is MOS$_{78}$-PBE, where the WTMAD-2\textsubscript{GMTKN55} is increased by $0.15$ \kcalmol.
The majority of this performance loss comes from the intermolecular non-covalent interactions subset, most of which are due to the S66 non-covalent interactions set.
During the optimization, the parameters {$a_{X}$} and $\omega$ values for MOS-PBEP86 are found to be unchanged.
For MOS-SCAN and MOS-PBE, only the optimized $\omega$ value changed.
On the other hand, for xMOS-PBEP86, both the {$a_{X}$} and $\omega$ values deviate from the unconstrained optimization.

\begin{figure*}[!htbp]
    \centering
    \includegraphics[width=\textwidth]{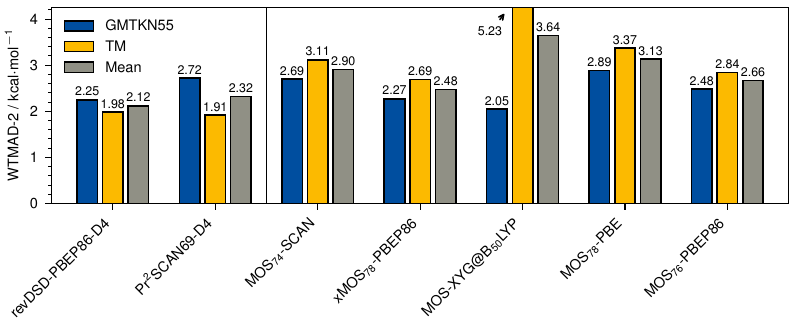}
    \caption{WTMAD-2 error statistics for GMTKN55, the transition-metal sets, and their mean for selected MOS-double-hybrids, revDSD-PBEP86-D4, and Pr\textsuperscript{2}SCAN69-D4. The results for revDSD-PBEP86-D4 and Pr\textsuperscript{2}SCAN69-D4 are extracted from reference.\cite{r2SCAN-D4-dh}}
    \label{fig:TM-GM-errors}
\end{figure*}

\subsection{Transition Metal Reactions}

The mean absolute errors of the present functionals on the nine transition metal subsets with their respective total WTMAD-2\textsubscript{TM} are presented in Tab.~\ref{tab:TMsubset_wtamd2} and \sitab{S23}.

For both, MOS$_{76}$-PBEP86	and xMOS$_{78}$-PBEP86 we find a very similar WTMAD-2\textsubscript{TM} of $2.84$ and $2.69$~\kcalmol, respectively.
This is also reflected in the mean absolute errors of individual subsets. 
MOS$_{76}$-PBEP86 is found to perform better on the ROST61 and WCCR10 sets, whereas xMOS$_{78}$-PBEP86 performs better on the TMIP and CUAGAU-2.

A comparison of both MOS double-hybrids to revDSD-PBEP86-D4 or Pr\textsuperscript{2}SCAN69-D4 reveals poor performance on nearly all transition-metal sets.\cite{r2SCAN-D4-dh}
Of those sets, the barrier-height sets are the least affected, which is likely due to the large amounts of Hartree--Fock exchange.
At the same time, this large amount is likely responsible for the deterioration of performance for the transition-metals.
Large amounts of Fock-exchange can sometimes be favorable for main-group thermochemistry -- as observed for GMTKN55 -- but can potentially be problematic for transition-metal chemistry where a higher degree of static correlation effects can be expected.
Especially open-shell transition-metal complexes are prone to static correlation effects as found on the ROST61, CUAGAU-2, and TMIP.\cite{Neugebauer_Vuong_Weber_Friesner_Shee_Hansen_2023}

While MOS-XYG@B$_{50}$LYP showed excellent accuracy on the GMTKN55 for main-group thermochemistry, it is not transferable to our TM datasets. It has the highest WTMAD-2\textsubscript{TM} of $5.23$~\kcalmol\ among all tested functionals.

Comparing the mean WTMAD-2 values of GMTKN55 and TM datasets (WTMAD-2\textsubscript{Mean}), it is evident that xMOS$_{78}$-PBEP86 is the best performer among the new MOS double hybrids. However, it still underperformed compared to regular SOS/SCS-MP2-based dispersion-corrected functionals, such as revDSD-PBEP86-D4 or Pr\textsuperscript{2}SCAN69-D4 (Fig.~\ref{fig:TM-GM-errors}).

\begin{table*}[!htbp]
\caption{WTMAD-2 and MAEs of metal-organic benchmark sets in \kcalmol. The revDSD-PBEP86-D4 and Pr\textsuperscript{2}SCAN69-D4 error statistics are taken from reference.\cite{r2SCAN-D4-dh}
}
\begin{ruledtabular}
\renewcommand{\arraystretch}{1.4}
\begin{tabular}{lcccccccccc}
\textbf{Functional}	&	\textbf{WTMAD-2\textsubscript{TM}}	&	\textbf{CUAGAU-2}	&	\textbf{LTMBH}	&	\textbf{MOBH35}	&	\textbf{MOR41}	&	\textbf{ROST61}	&	\textbf{TMBH}	&	\textbf{TMCONF16}	&	\textbf{TMIP}	&	\textbf{WCCR10}	\\
\hline
MOS$_{76}$-PBEP86	&	2.84	&	4.17	&	0.88	&	1.65	&	3.82	&	3.43	&	1.13	&	0.22	&	14.85	&	1.65	\\
xMOS$_{78}$-PBEP86	&	2.69	&	4.01	&	0.80	&	1.62	&	3.79	&	2.95	&	1.13	&	0.23	&	10.31	&	1.80	\\
MOS-XYG@B$_{50}$LYP	&	5.23	&	6.74	&	1.54	&	4.07	&	5.86	&	5.97	&	3.10	&	0.21	&	11.47	&	5.19	\\
\hline
revDSD-PBEP86-D4\cite{revdsdpbep86d4,r2SCAN-D4-dh}	&	1.98	&	2.89	&	0.57	&	1.42	&	2.81	&	2.04	&	0.84	&	0.18	&	9.01	&	1.45	\\
Pr\textsuperscript{2}SCAN69-D4\cite{r2SCAN-D4-dh}	&	1.91	&	3.23	&	0.40	&	1.62	&	2.40	&	1.96	&	0.98	&	0.16	&	7.96	&	1.86	\\
\end{tabular}
\end{ruledtabular}
\label{tab:TMsubset_wtamd2}
\end{table*}

\section{\label{sec:conclusions}Conclusions}

We have proposed a new variety of double hybrid functionals using the modified opposite-spin MP2 as the nonlocal correlation component.
From our investigation of MOS-DHs and their respective DOD- and DSD-family counterparts with the aid of the GMTKN55 and nine additional transition-metal data sets, we can conclude the following:

\begin{enumerate}[label=(\alph*)]
    \item Except for the MOS-XYG functional, the MOS-MP2-based double hybrids always outperform their SCS-MP2-based counterparts on the GMTKN55 without the addition of any dispersion correction.
    
    \item The major benefit of distance-dependent MP2 compared to regular MP2 in double hybrids originates from the poor description of long-range correlation effects, which is especially important for inter- and intra-molecular non-covalent interactions.
    
    \item For a fixed value of HF-exchange in MOS-DHs, gradually increasing attenuation parameter ($\omega$) requires a systematically smaller amount of total MOS-MP2 correlation.
    
    \item Among all the MOS-DHs tested in the present study, the six-parameter MOS-XYG@B$_{50}$LYP has the lowest WTMAD-2\textsubscript{GMTKN55} of 2.05 \kcalmol, but with a large percentage of HF-exchange and large a $\omega$ MOS-MP2 attenuation parameter.
    However, for the transition-metal-involving reactions, it yields the worst WTMAD-2\textsubscript{TM}.
    
    \item The four-parameter xMOS$_{78}$-PBEP86 shows the best overall performance, balancing main-group and transition-metal chemistry performance.
    However, in terms of the WTMAD-2\textsubscript{Mean}, the dispersion uncorrected xMOS$_{78}$-PBEP86 performs worse than Pr\textsuperscript{2}SCAN69-D4 and revDSD-PBEP86-D4.    
    
    \item When D4 correction is incorporated into double hybrids, using the MOS-MP2 scheme does not provide additional benefits compared to simple SOS-MP2.  
\end{enumerate}

\section*{Acknowledgement}

G.S. thanks Prof. Jan M. L. Martin (Dept. of Molecular Chemistry and Materials Science, Weizmann Institute of Science, Reḥovot, Israel) for kindly allowing access to the ChemFarm HPC cluster of the Weizmann Institute Faculty of Chemistry. All calculations for parameterizing the modified opposite-spin-scaled double-hybrid functionals were performed on ChemFarm.

L.W. greatly acknowledges the support of the Stiftung Stipendien-Fonds des Verbandes der Chemischen Industrie e.V. through its Kekul\'{e} Fellowship.

L.W. would like to extend his gratitude to Prof. Stefan Grimme for his exceptional support and for granting access to the computational resources at the Mulliken Center for Theoretical Chemistry.

\section*{Data availability statement}

Raw data can be obtained upon reasonable request.

\section*{References}
\bibliography{MOS-DH}%

\end{document}